\documentclass{PoS}

\usepackage{amsmath,amsfonts,mathtools,amssymb,slashed}
\usepackage[export]{adjustbox}

\title{Freeze-in leptogenesis with 3 right-handed neutrinos}

\ShortTitle{Freeze-in leptogenesis with 3 right-handed neutrinos}

\author{\speaker{Michele Lucente},$^{a}$\thanks{This project has received funding from the European Union's Horizon 2020 research and innovation programme under the Marie Sklodowska-Curie grant agreement No 750627.}\phantom{o} Asmaa Abada,$^{b}$ Giorgio Arcadi,$^{c}$ Valerie Domcke,$^{d}$ Marco Drewes$^{a}$ and Juraj Klaric$^{e,f}$
    \\
    \\
\llap{$^{a}$}Centre for Cosmology, Particle Physics and Phenomenology - CP3, Universit\'{e} catholique de Louvain, Chemin du Cyclotron 2, 1348 Louvain-la-Neuve, Belgium\\
\llap{$^{b}$}Laboratoire de Physique Th\'eorique (UMR8627), CNRS, Univ. Paris-Sud, Universit\'e Paris-Saclay, 91405 Orsay, France\\
\llap{$^{c}$}Max-Planck-Institut f\"ur Kernphysik (MPIK), 69117 Heidelberg, Germany\\
\llap{$^{d}$}Deutsches Elektronen-Synchrotron (DESY), 22607 Hamburg, Germany\\
\llap{$^{e}$}Technische Universit\"at M\"unchen (TUM), Garching, Germany\\
\llap{$^{f}$}Institute of Physics, Laboratory for Particle Physics and Cosmology (LPPC), Ecole Polytechnique F\'{e}d\'{e}rale de Lausanne (EPFL), CH-1015 Lausanne, Switzerland\\ \\
        E-mail: \email{michele.lucente@uclouvain.be}, \email{asmaa.abada@th.u-psud.fr}, \email{arcadi@mpi-hd.mpg.de}, \email{valerie.domcke@desy.de}, \email{marco.drewes@uclouvain.be}, \email{juraj.klaric@epfl.ch}}

\abstract{We provide the first systematic study of the viable parameter space for leptogenesis in the type-I seesaw model with three right-handed neutrinos whose Majorana masses lie below the electroweak scale.
We highlight the very rich phenomenology of this scenario and discuss several mechanisms that can help to enhance the baryon asymmetry of the Universe.
This allows for much larger heavy neutrino mixing angles than in the minimal scenario with two heavy neutrinos, resulting in solutions that can be probed with the LHC and other experiments.
The light neutrino masses can be protected from radiative corrections by an approximate symmetry related to generalised lepton number.
}

\FullConference{The 39th International Conference on High Energy Physics (ICHEP2018)\\
		4-11 July, 2018\\
		Seoul, Korea}

\begin{document}

\section{Introduction}
The existence of new physics beyond the Standard Model (SM) of particle physics is called for by several experimental results and theoretical arguments. Among these, there are two robust and seemingly unrelated observations: the fact that neutrinos are massive and leptons mix, and the measured baryon asymmetry of the Universe (BAU).
By looking at the SM field content, on the other hand, one immediately notices a certain asymmetry in it: while all the fermionic fields exist in both states of chirality (left- and right-handed) this is not the case for neutrinos, that are present only with their left-handed components. It is thus natural to ask what happens if the SM field content is extended to include right-handed neutrinos as well, and study the phenomenological implications of this extension. In this work we show that the SM extended with three right-handed neutrinos 
with masses below the electroweak scale
can provide a simultaneous explanation for both neutrino masses (and lepton mixing) and the BAU, as observed, and that a relevant fraction of the resulting parameter space can be tested by already running experiments, including ATLAS, CMS, LHCb, NA62, T2K and Belle II, cf.~\cite{Abada:2018oly} for further details.

\section{Neutrino masses and leptogenesis}
The addition of a number $n$ of right-handed neutrinos $\nu_{R}$ to the SM field content implies the appearance of the following operators in the most general renormalizable Lagrangian,
\begin{equation}
    \mathcal{L}
  = \mathcal{L}_\text{SM} +\left( \frac{\operatorname i}{2} \overline{\nu_{R i}} \slashed \partial \nu_{R i}
 - F_{a i} \overline{\ell_L}_a \varepsilon \phi^* \nu_{R i}
 - \frac{1}{2} \overline{\nu_{R i}^c} \left(M_M\right)_{i j} \nu_{R j}
 + \text{h.c.}\right), \label{eq:Lagrangian}
\end{equation}
where $i,j=1,\dots,n$ denote the right-handed neutrino flavours, $\ell_{L a}$ are the SM lepton doublets (with $a=e,\mu,\tau$), $\phi$ is the Higgs doublet and $\varepsilon$ the totally antisymmetric 2-dimensional tensor; we have suppressed the $SU(2)$ indices. The new parameters in the Lagrangian are the dimensionless Yukawa couplings $F_{a i}$ and the Majorana mass terms $\left(M_M\right)_{i j}$. After the Higgs field acquires a vacuum expectation value, $\left< \phi \right> = v$ (with $v=174$ GeV at temperature $T=0$), the Lagrangian~(\ref{eq:Lagrangian}) generates a non-vanishing neutrino mass matrix $m_\nu$,
\begin{equation}
    m_\nu = m_\nu^{\rm tree} + \delta m_\nu^{\rm{1-loop}}  = - v^2 F M_M^{-1} F^T + F M_N^{\rm diag} l(M_N^{\rm diag})F^T,
\label{seesaw}
\end{equation}
where $M_N^{\rm diag}$ is a diagonal matrix containing the physical heavy-neutrino masses and $l(M_N^{\rm diag})$ is a loop function~\cite{Pilaftsis:1991ug}. It is remarkable that the very same Lagrangian~(\ref{eq:Lagrangian}) contains the necessary ingredients for baryogenesis as well: the Yukawa couplings $F_{ai}$ are in general complex numbers, providing a new source of $CP$-violation; baryon number is violated in the SM by electroweak sphaleron processes, while the new degrees of freedom $\nu_R$ can deviate from thermal equilibrium during their evolution in the early Universe.

\subsection{Leptogenesis realisations}
The temperatures at which the heavy neutrinos enter and subsequently deviate from thermal equilibrium allow to classify two main scenarios for leptogenesis. In the first one, usually dubbed \emph{thermal} or \emph{freeze-out leptogenesis}~\cite{Fukugita:1986hr}, heavy neutrinos have Yukawa couplings sizeable enough such that they equilibrate well before the electroweak phase transition; due to 
cosmic
expansion they decouple afterwards, freeze-out and eventually decay out of equilibrium. It is this out-of-equilibrium decay, 
together with the CP-violating nature of the Yukawa couplings $F_{ai}$, 
that generates a lepton asymmetry, which is subsequently converted by sphalerons into a non-vanishing baryon asymmetry. This realisation is theoretically very attractive, but is difficult to test in laboratory experiments, since it requires heavy neutrino masses larger than $\mathcal{O}(10^6 \text{ GeV})$ in the case of a non-degenerate mass spectrum~\cite{Moffat:2018wke}, and in any case above the electroweak scale if the heavy neutrinos feature a degenerate mass spectrum (resonant leptogenesis)~\cite{Pilaftsis:2003gt}.
In the second scenario, 
usually dubbed \emph{low-scale} or \emph{freeze-in leptogenesis}~\cite{ARS_AS}, 
right-handed neutrinos have smaller Yukawa couplings, 
such that they equilibrate around of after the electroweak phase transition temperature. The generation of a lepton asymmetry in this case proceeds during the production of right-handed neutrinos from the thermal plasma, while they deviate from the thermal equilibrium abundance, and is mediated by CP-violating lepton-number conserving and lepton-number violating processes. The former include the generation of heavy neutrinos from the thermal bath and their flavour oscillations, 
while the latter processes (in particular Higgs decays) are relevant at later times. 

\subsection{Testability and fine-tuning}
From the seesaw relation in Eq.~(\ref{seesaw})
it is evident that, in order to reproduce the observed neutrino masses, low-scale leptogenesis is realised by heavy neutrinos with masses at the GeV scale. This mass scale is accessible in current experiments, and such scenario is thus in principle testable.
However, the testability of a parameter point is determined by both the masses of the new particles and their couplings with the SM, proportional to the active-sterile mixings $U_{a i}$.
In the absence of any structure in the $F$ and $M_M$ matrices, the naive seesaw scaling $m_\nu \simeq - v^2 F_{ai}^2 / M_M $ predicts an upper bound for the mixing, $\left| U_{ai} \right| \simeq \sqrt{m_\nu /M_M} \lesssim 10^{-5} \sqrt{\text{GeV}/M_M}$, a value which is orders of magnitude below the current experimental bounds. Eq.~(\ref{seesaw}) is however a complex matrix equation, and cancellations are possible: these can make the observed neutrino mass scale compatible with Yukawa couplings (and thus active-sterile mixings) sizeably larger than the values predicted by the naive seesaw scaling. Such cancellations can result from an underlying symmetry, a notable example of which is an approximate lepton number conservation: light (Majorana) neutrino masses violate lepton number, and are thus necessarily suppressed if such a symmetry is present in the Lagrangian, while lepton-number conserving processes mediated by heavy neutrinos (actively looked for in laboratory experiments) are not suppressed by it. Alternatively, accidental cancellations are possible, making testable mixing values compatible with the observed neutrino mass scale. These two possibilities are nonetheless very different from a theoretical viewpoint: if a symmetry is present in the Lagrangian, it will be manifest at any order in perturbation theory, making neutrino masses stable under radiative corrections, while it is extremely unlikely that accidental cancellations appear both at tree and loop-level. In our analysis we do not impose any symmetry, but quantify a posteriori the level of fine tuning for each solution, by defining the function $ f.t. (m_\nu) = \sqrt{\sum_{i=1}^3 \left( \left( m_i^\text{loop}-m_i^\text{tree} \right) /m_i^\text{loop}\right)^2} $, where $m_i^\text{tree}$ and $m_i^\text{loop}$ are the light neutrino masses computed at tree and 1-loop level, respectively. The smaller is $f.t. (m_\nu)$, the more neutrino masses are stable under radiative corrections, suggesting the presence of underlying symmetries if the active-sterile mixing is sizeably larger than the naive seesaw scaling.

\section{Results and conclusion}
\begin{figure}
\centering
\vspace{-0.1cm}
\adjincludegraphics[width=0.49\textwidth,trim={0 0 0 0.65cm},clip]{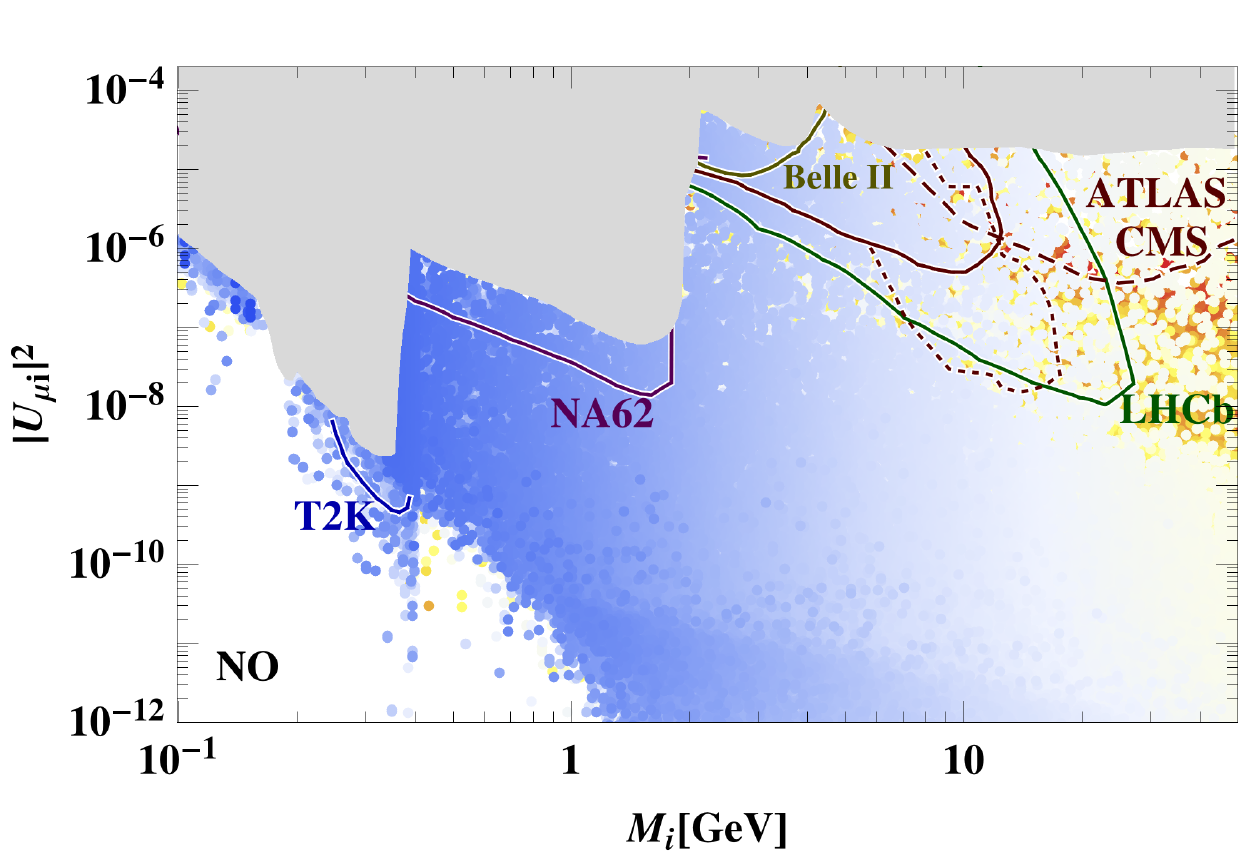}
\hspace{-0.3cm} \vspace{-0.35cm}
\adjincludegraphics[width=0.5\textwidth,trim={0 0 0 0.65cm},clip]{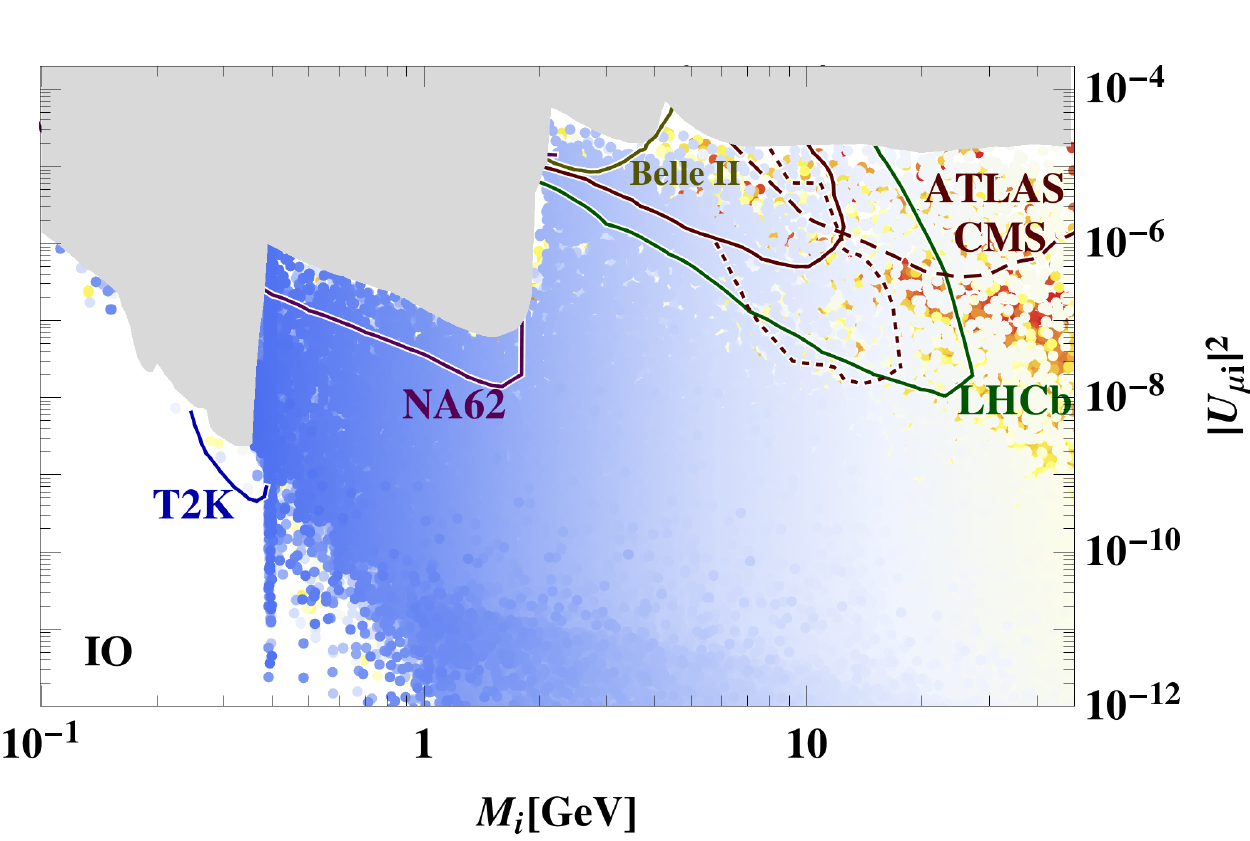}
\includegraphics[width=0.3\textwidth]{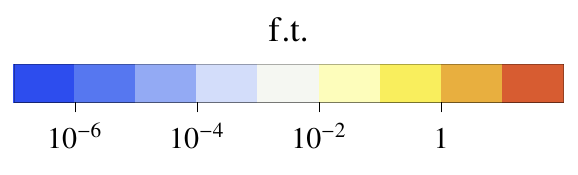}
\vspace{-0.3cm}
\caption{Active-sterile mixing in the muon flavour for the viable BAU solutions as a function of the heavy  neutrino mass, for a normal (left) and inverted (right) ordering in the light neutrino mass spectrum. The grey region is experimentally excluded, while the lines show the expected sensitivities for ongoing experiments.}
\label{fig:mass_mixing}
\end{figure}

We report in Fig.~\ref{fig:mass_mixing} the active-sterile mixing in the muon flavour as a function of the heavy neutrino mass, for the viable solutions reproducing both active neutrino parameters and the BAU. Similar results are obtained for the other flavours~\cite{Abada:2018oly}.
It is evident from Fig.~\ref{fig:mass_mixing} that solutions with large mixings (up to the current experimental bounds) are viable in the considered mass range $[0.1, 50]$ GeV, and no fine-tuning is required despite of the large value of the active-sterile mixing. 
This is possible due to a number of new effects, including lepton number violating oscillations and decays, flavour asymmetric washouts and a resonant enhancement due to matter effects (similar to the Mikheyev-Smirnov-Wolfenstein effect), cf.~\cite{Abada:2018oly} for a detailed discussion.
In conclusion, the freeze-in leptogenesis scenario with three right-handed neutrinos provides a viable solution to the BAU, compatible with constraints from neutrino physics, and exhibits a very rich phenomenology, allowing for (not fine-tuned) solutions with comparably large heavy neutrino mixing angles that can be probed at the LHC and can also give a sizeable contribution to neutrinoless double $\beta$ decay~\cite{Abada:2018oly}.


\begin{thebibliography}{99}
\bibitem{Abada:2018oly}
  A.~Abada, G.~Arcadi, V.~Domcke, M.~Drewes, J.~Klaric and M.~Lucente, \emph{Low-scale leptogenesis with three heavy neutrinos},
  arXiv:1810.12463 [hep-ph].
  
\bibitem{Pilaftsis:1991ug}
  A.~Pilaftsis,
  \emph{Radiatively induced neutrino masses and large Higgs neutrino couplings in the standard model with Majorana fields},
  Z.\ Phys.\ C {\bf 55} (1992) 275
  [hep-ph/9901206].

\bibitem{Fukugita:1986hr}
  M.~Fukugita and T.~Yanagida,
  \emph{Baryogenesis Without Grand Unification},
  Phys.\ Lett.\ B {\bf 174} (1986) 45.
  
\bibitem{Moffat:2018wke}
  K.~Moffat, S.~Pascoli, S.~T.~Petcov, H.~Schulz and J.~Turner,
  \emph{Three-flavored nonresonant leptogenesis at intermediate scales},
  Phys.\ Rev.\ D {\bf 98} (2018) no.1,  015036
  [arXiv:1804.05066 [hep-ph]].
  
\bibitem{Pilaftsis:2003gt}
  A.~Pilaftsis and T.~E.~J.~Underwood,
  \emph{Resonant leptogenesis},
  Nucl.\ Phys.\ B {\bf 692} (2004) 303
  [hep-ph/0309342].
  
  \bibitem{ARS_AS}
  E.~K.~Akhmedov, V.~A.~Rubakov and A.~Y.~Smirnov,
  \emph{Baryogenesis via neutrino oscillations},
  Phys.\ Rev.\ Lett.\  {\bf 81} (1998) 1359
  [hep-ph/9803255];
  T.~Asaka and M.~Shaposhnikov,
  \emph{The nuMSM, dark matter and baryon asymmetry of the universe},
  Phys.\ Lett.\ B {\bf 620} (2005) 17
  [hep-ph/0505013].
\end{thebibliography}
\end{document}